\documentclass[11pt]{article}
\usepackage{amsmath}
\usepackage{epsfig,bm}
\def\b{\begin}
\def\e{\end}
\def\t{thebibliography}
\def\bi{\bibitem}
\def\be{\b{equation}}
\def\ee{\e{equation}}

\def\nt{\noindent}

\begin{document}

\title{AXIOMATIC FOUNDATIONS FOR THE PRINCIPLE OF ENTROPY INCREASE }

\author{Qi-Ren Zhang
\\Department of Technical Physics, Peking University , Beijing,100871,
China}

\maketitle

\vskip0.3cm

\begin{abstract}
We show that the principle of entropy increase may be exactly
founded on a few axioms valid not only for quantum and classical
statistics, but also for a wide range of statistical processes.
\end{abstract}

\bigskip

\section{Introduction}The second law of thermodynamics, or the principle of entropy
increase, is an exact law of nature. To explore its foundation is
one of the most important topics in physics for more than a century.
From the daily life we know, matter always approaches its
equilibrium state. Text books\cite{m}-\cite{z} tell us that the
equilibrium state is a state with maximum entropy, and the state of
a macroscopic system with larger entropy is more probable. It almost
explains the above daily life experience. However, whether the
system always goes from a less probable state to a more probable
state, or why the principle of entropy increase works, is still an
open question. The H-theorem of Boltzmann is a classical proof for
definite approaching to equilibrium. It is based on a model of
colliding classical particle system for the macroscopic matter,
therefore is not general enough, even from the view point of the
classical statistical physics. Recently we gave an exact and general
proof for the principle of entropy increase by use of general
principles of quantum theory\cite{z1}. In this paper, we would put
the principle on an axiomatic basis, consistent with classical and
quantum mechanics, but not special for them. We collect definitions
and axioms in section 2, lemmas and theorems in section 3.  Among
them, theorem 4 is exactly the principle of entropy increase.
Section 4 is a discussion.

\section{Definitions and axioms}
{\bf Definition 1}: The evolution is a process not interrupted by
measurement.

\noindent {\bf Definition 2}: A system is a collection of objects,
its evolution in time is determined by itself, and is not correlated
with any other object.

\noindent{\bf Definition 3}: The state of a system at a given time
is a property of the system at that time, which determines which
observables are certain, what are their value, as well as the change
of this property itself at that time.

It means the existence of a state is conditional. If there is a
state for the system under consideration at a time $t_0$, it must
have a state at any other time $t$ in the course of evolution which
is determined by the original state at the time $t_0$. The
differential equation governing the state evolution is of the first
order in time, which in turn means that two different states keep
different during the evolution. These points are true both for
classical and quantum mechanics. On a set of independent events one
may define probability distribution.

\noindent {\bf Definition 4}: Independence between two states means:
if the system stays in one of them  the observer can definitely not
see the property which defines another state.

\noindent{\bf Definition 5}: A set of states for a system is
complete if and only if any other state of the system depends on at
least one state in it.

\noindent In classical mechanics, two different states are always
independent from each other; while in quantum mechanics,
independence of states means they are orthogonal to each other. Two
different but nonorthogonal states depend on each other in the
sense, that there is a nonzero probability $W_{ab}$ to see the
property of one state $a$ while the system stays in another state
$b$. In classical mechanics $W_{ab}=\delta_{ab}$, while in quantum
mechanics the relation is relaxed to be $W_{ab}=W_{ba}$. Both in
classical and in quantum mechanics, two independent states keep
independent from each other through their evolution. Since an
orbital state of a particle occupies a finite volume $h^3$ in its
phase space, quantum state is numerable. Classical state is
innumerable in its original form. However, it is an effective method
in classical statistical mechanics, to let the phase volume of a
state finite, so that make the state numerable, and let the phase
volume approaches zero at the final stage of derivation. If
classical theory is applicable to the problem, this method always
gives right answer.

\noindent {\bf Definition 6}: The state with a certain value of a
given observable is called an eigenstate of this observable, the
corresponding value of the observable is called its eigenvalue.

\noindent In classical mechanics, every state is an eigenstate of
all observables. But in quantum mechanics, eigenstate of an
observable has to be solved from the eigenequation of the
observable. However, in both cases a set of all independent
eigenstates for a given observable is complete.

\noindent {\bf Definition 7}: If a set of observables may be
measured simultaneously with certain outcome,  and the result is
complete enough to determine the state of the measured system, the
set is called a complete set of observables for the system.

\noindent {\bf Definition 8}: Measurement of a complete set of
observables on a system is called a complete measurement.

While a complete measurement determines the state of the system, an
incomplete measurement cannot determine the state but may determine
a probability distribution of the system over a set of different
states. In classical mechanics, it means a probability distribution
over a complete set of independent states. In quantum mechanics one
handles this situation by a Hermitian operator, called the density
operator. Its eigenstates are independent of each other, and the
eigenvalues may be regarded as probabilities of finding that the
system stays in the corresponding states respectively. In both
cases, an incomplete measurement determines a probability
distribution of the system over a complete set of independent
states. Denoting the $n$th state of the set by $|n\rangle$ and the
probability of the system staying in the state $|n\rangle$ by a
non-negative number $W_n$, we have the normalization relation
$\sum_nW_n=1$, and

\noindent {\bf Definition 9}:  The information of a system is given
by \be {\cal I}\equiv\sum_nW_n\ln W_n\;.\label{1}\ee The summation
is over the complete set of independent states, and the information
means the amount of information for short.

In classical and quantum physics, it is always possible to divide a
macroscopic system into subsystems, so that every subsystem is still
large enough to be macroscopic but is already macroscopically
uniform, and the microscopic non-uniformity is still negligible.

\noindent {\bf Definition 10}: The uniform system is a system, in
which a kind of observable (intensive observable) takes the same
value everywhere, and the values of other kinds of observables
(extensive observables) are proportional to each other.

\noindent {\bf Definition 11}: The divisible system is either a
uniform system or a system which may be divided into uniform
subsystems.

\noindent For a divisible system one may define the entropy.

\noindent {\bf Definition 12}:  In unit of Boltzmann constant $k$
the entropy is given by \be S=-\sum_i{\cal I}_i\; ,\label{2}\ee in
which subscript $i$ specifies its subsystems all being uniform. It
is the negative sum of the information of these subsystems.

We assume our system satisfies the following axioms:

\noindent {\bf Axiom 1}: The system is in one of its states at a
given time.

\noindent {\bf Axiom 2}: The probability $W_{ab}$ of finding the
property of state $a$ in the state $b$ equals the probability
$W_{ba}$ of finding the property of state $b$ in the state $a$.

\noindent {\bf Axiom 3}: Two eigenstates of the same observable with
different eigenvalues are independent from each other.

\noindent {\bf Axiom 4}: The set of all independent eigenstates for
a given observable is complete.

\noindent {\bf Axiom 5}: The set of independent states for a system
is numerable.

\noindent {\bf Axiom 6}: The state at a time determines the state at
any other time for the same system throughout the whole process of
evolution.

\noindent {\bf Axiom 7}: Independent states keep independent from
each other during the evolution in time.

\noindent {\bf Axiom 8}: The state of the system after a complete
measurement is the one determined by the outcome of the measurement.

\noindent {\bf Axiom 9}: The system may be described by a
probability distribution over a complete set of independent states.

\noindent {\bf Axiom 10}: The system is divisible.

In classical and quantum statistical physics, these axioms are
exactly satisfied. We expect they may still be satisfied in future
new physics, and also be satisfied in some processes other than
those in physics. It would make the results derived from them not
only exactly true in physics but also applicable to a wide range of
other problems

\section{ Lemmas and theorems}
Now let us remind you some mathematical inequalities. One can find
them and their proofs elsewhere\cite{z1,e}. Mathematically, we
define $0\ln0\equiv\lim_{\xi\rightarrow 0}(\xi\ln\xi)=0$.

\nt {\bf Lemma 1.} For any non-negative number $x$ we have  \be x\ln
x\geq x-1\; , \label{11}\ee the equality holds when and only when
$x=1$.

\nt {\bf Lemma 2.} For sets $[w_i]$ and $[x_i]$ of non-negative
numbers with $\sum_ix_i=1$, we have \be
\sum_ix_iw_i\ln\sum_{i^\prime}x_{i^\prime}
w_{i^\prime}\leq\sum_ix_iw_i\ln w_i\; .\label{12}\ee

\nt {\bf Lemma 3.} For sets $[W_i]$ and $[T_{ij}]$ of non-negative
numbers with \be \sum_iW_i=1\;\;\;\;\; \mbox{and}\;\;\;\; \;
\sum_iT_{ij}=\sum_jT_{ij}=1 \; ,\label{0}\ee we have \be W_j^\prime
\equiv\sum_iW_iT_{ij}\geq 0 \;\;\;\;\mbox{for every} \;j
,\label{a}\ee\be\sum_jW_j^\prime=1\; ,\label{b}\ee and \be
\sum_jW_j^\prime\ln W_j^\prime\leq\sum_iW_i\ln W_i\; .\label{13}\ee

\nt {\bf Lemma 4.} For positive numbers $[W_{ij}]$,
$W_i=\sum_jW_{ij}$ and $W_j^\prime=\sum_iW_{ij}$, with
$\sum_{ij}W_{ij}=1$, we have \be
\sum_iW_i=1\;,\;\;\;\;\;\;\sum_jW_j^\prime =1\;,\label{14}\ee and
\be \sum_{ij}W_{ij}\ln W_{ij}\geq \sum_iW_i\ln W_i
+\sum_jW_j^\prime\ln W_j^\prime\; .\label{15}\ee The equality holds
when and only when $W_{ij}=W_iW_j^\prime$ for all $ij$, it is that
the $W_{ij}$ may be factorized.

Consider a system. At time $t_0$, we do not know its state, but know
the probability distribution over a complete set of its independent
states. Its information is therefore given by (\ref{1}). We have

\noindent{\bf Theorem 1 (information conservation)}: The information
of a system does not change in the course of evolution.

{\bf Proof:} From the definition 9 we see, the information of a
system relates only to the probability distribution over a complete
set of its independent states, irrelevant to the contents of these
states. Since, according to the axiom 7, the independence of states
does not change in the evolution, the probability distribution and
therefore the information of the system does not change either. It
is \be {\cal I}(t)={\cal I}(t_0)\; ,\label{aa}\ee in which ${\cal
I}(t)$ and ${\cal I}(t_0)$ are information of the system at two
different time $t$ and $t_0$ respectively, in its course of
evolution. The theorem is proven.

According the axioms 3 and 4, a complete set $[L]$ of observables
for the system has a complete set of independent eigenstates.
Denoting this set by $[|m\rangle]$, and the probability of finding
the property of state $|m\rangle$ by $W^\prime_m$, the information
of the complete set  $[L]$ of observables for the system is defined
by \be {\cal I}_{[L]}\equiv\sum_mW^\prime_m\ln W^\prime_m\; .\ee
Denoting the probability of finding the property of state
$|m\rangle$ in the state $|n\rangle$ by $W_{m,n}$, we have  \be
W^\prime_m=\sum_n W_{m,n}W_n \; , \ee  \be \sum_m W^\prime_m =1 \;
,\ee and by axiom 2 also \be \sum_nW_{m,n}=\sum_mW_{m,n}=1\; .\ee
Since probabilities $W_{m,n}$ are non-negative, according to lemma 3
and equation (\ref{1}) we obtain \be {\cal I}_{[L]}\leq {\cal I}\; ,
\ee and therefore have proven

\noindent {\bf Theorem 2}: The information of a given complete set
of observables for the system is not more than the information of
the system itself.

Now, let us divide the system into two parts $a$ and $b$. Suppose
$[L_i]$, with $i=a$ or $b$, is a complete set of observables of part
$i$, $|n_i\rangle$ is their $n_i$th eigenstate, and $[|n_i\rangle]$
is a complete set of independent states for part $i$. Therefore
$[L_a,L_b]$ is a complete set of observables, and
$[|n_an_b\rangle]\equiv[|n_a\rangle|n_b\rangle]$ is a complete set
of independent states, both for the system. Denoting the probability
of finding the property of state $|n_an_b\rangle$ in the state
$|n\rangle$ by $W_{n_an_b,n}$, the probability of finding part $a$
in the state $|n_a\rangle$ and part $b$ in the state $|n_b\rangle$
is \be W_{n_an_b}=\sum_nW_{n_an_b,n} W_n\; ,\ee with normalization
\be \sum_{n_an_b}W_{n_an_b}=1\; .\label{z}\ee According to the
theorem 2, the information of observables $[L_a,L_b]$ for the system
is \be {\cal I}_{L_a,L_b}=\sum_{n_a,n_b}W_{n_an_b}\ln W_{n_an_b}\leq
{\cal I}\; .\label{f}\ee The probability of finding part $a$ in the
state $|n_a\rangle$ and the probability of finding the part $b$ in
the state $|n_b\rangle$ are \be
W_{n_a}=\sum_{n_b}W_{n_an_b}\;\;\;\;\mbox{and}\;\;\;\;W_{n_b}
=\sum_{n_a}W_{n_an_b}\; .\label{g}\ee respectively. In
(\ref{z}-\ref{g}), it is understood that the summation is over those
$n_a$ and $n_b$ only, for which $W_{n_an_b}>0$. According to axiom
8, after the measurement of $[L_i]$, the state of part $i$ would be
one in the set $[|n_i\rangle]$. The probability for the presence of
state $|n_i\rangle$ is $W_{n_i}$. The information for part $i$ is
\be {\cal I}_i=\sum_{n_i}W_{n_i}\ln W_{n_i}\; .\label{gg}\ee From
lemma 4 and equations (\ref{f}-\ref{gg}) we see \be {\cal I}_a+{\cal
I}_b\leq {\cal I}\; .\label{h}\ee The equality holds when and only
when $W_{n_an_b}=W_{n_a}W_{n_b}$ for all $n_a$ and $n_b$, it is that
the probability distribution is factorized. The later means two
parts of the system do not correlate with each other. We may further
subdivide the parts and apply (\ref{h}) to them again and again, the
result is the statement\be \sum_i{\cal I}_i\leq {\cal I}\;
,\label{ii}\ee in which the summation is over all parts of the
system. Therefore we have proven

\noindent {\bf Theorem 3}: The sum of information of all parts of
the system is no more than the information of the system itself.

According to the axiom 10, we may make every part of the system be
uniform. In this case, by the definition 12 for the entropy and the
equation (\ref{ii}), we see \be S\geq -{\cal I}\label{ij}\ee for a
system. The equality holds when and only when parts of the system
are not correlated to each other. We then arrive at

\nt {\bf Theorem 4 (principle of entropy increase)}: The entropy of
a system if changes can only increase.

{\bf Proof:} We would prove the theorem operationally. Suppose we
measured the entropy $S(t_0)$ of the system at the beginning time
$t_0$. According to the definition 12, it means that we measured the
entropy of every uniform part of the system and summed them up. This
operation had to destroy the correlation between these parts, and
made the probability distribution of the system factorized to a
product of probability distributions of its parts. By eq.(\ref{ij})
and the discussion after it, we see \be S(t_0)=-{\cal I}(t_0)\;
,\label{kk}\ee in which ${\cal I}(t_0)$ is the information of the
system at time $t_0$ after the measurement. After this first
measurement the system evolves according to its own dynamics with
information conservation (theorem 1). In this course various parts
of the system become correlated because the interaction between
them. The probability distribution of the system will not keep being
factorized to the product of probability distributions of its parts.
Let us measure the entropy $S(t)$ at the time $t>t_0$ in the course,
by eqs.(\ref{ij}), (\ref{aa}) and (\ref{kk}), we see \be S(t)\geq
-{\cal I}(t)=-{\cal I}(t_0)=S(t_0)\; , \label{bb}\ee in which ${\cal
I}(t)$ is the information of the system just before the measurement
at time $t$. If the evolution of the system is interrupted by
measurements at times $t_0<t_1<t_2<...<t_{n-1}<t$, by the arguments
resulting in (\ref{bb}) we see   $S(t)\geq S(t_{n-1}\geq ...\geq
S(t_2)\geq S(t_1)\geq S(t_0)$. Anyway, we have \be S(t)\geq S(t_0)\;
.\ee The interaction between different parts of the system makes
these parts be correlated, which in turn makes the entropy of the
system strictly increase. The theorem is therefore proven.

\section{Discussion}
The proof here is quite general. Beside the axioms stated in section
2, nothing is assumed. Both classical and quantum statistics satisfy
these axioms. Therefore, the principle of entropy increase is
exactly true in them. The axioms are not too special, and may also
be satisfied by a wide class of statistical processes. It means,
perhaps we can find some statistical science other than physics, in
which the principle of entropy increase is applicable as well.

From the proof we learn that the entropy of a system increases only
because that, when one considers it he always neglects the
correlation information between different parts of the system. It
emphasizes the importance of the correlation information in a
complete statistical science.

This work is supported by the National Nature Science Foundation of
China with Grant number 10305001.

\b{\t}{99} \bi{m} A. Munster, Statistical Thermodynamics (Academic
Press, Berlin, 1969-74) \bi{l} L. D. Landau and E. M. Lifshitz,
Statistical Physics, 3rd edition (Butterworth Heinemann, Oxford,
1980) \bi{g} W. Greiner, L. Neise, and H. St\"{o}cker ,
Thermodynamics and Statistical Mechanics (Springer-Verlag, New York,
Inc. , 1995) \bi{z} Qi-Ren Zhang, Statistical Mechanics (in Chinese)
(Science Press, Beijing 2004)\bi{z1} Q.R.Zhang   quant-ph/0610005
\bi{e}H. Everett, The theory of universal wave function in the
many-worlds interpretation of quantum mechanics, B.S.De Witt and N.
Graham eds. (Princeton University press, Princeton, 1973) \e{\t}
\end{document}